\documentclass[twocolumn,showpacs,preprintnumbers,
               superscriptaddress]{revtex4}
\usepackage{graphicx,epsf}
\newcommand{\bfk}{\mbox{{\boldmath $k$}}}
\newcommand{\bfp}{\mbox{{\boldmath $p$}}}
\newcommand{\bfq}{\mbox{{\boldmath $q$}}}
\newcommand{\bfgamma}{\mbox{{\boldmath $\gamma$}}}
\newcommand{\Tr}{{\rm Tr}}

\begin{document}

\title{Non-Fermi Liquid Behavior Induced by Resonant Diquark-pair 
Scattering in Heated Quark Matter
}
\author{Masakiyo Kitazawa}
\affiliation{Yukawa Institute for Theoretical Physics,
Kyoto University, Kyoto 606-8502, Japan}
\affiliation{Department of Physics,
 Kyoto University, Kyoto 606-8502, Japan}
\author{Teiji Kunihiro}
\affiliation{Yukawa Institute for Theoretical Physics,
Kyoto University, Kyoto 606-8502, Japan}
\author{Yukio Nemoto}
\affiliation{Department of Physics, 
Nagoya University, Nagoya, 464-8602 Japan}

\begin{abstract}
We show how the quasiparticle picture of quarks changes 
near but above the critical temperature $T_c$
of the color-superconducting phase transition 
in the heated quark matter.
We demonstrate that a non-Fermi liquid behavior of 
the matter 
develops drastically when the diquark 
coupling constant is increased
owing to the coupling of the quark with the pairing soft mode: 
We clarify that the depression and eventually
the appearance of a gap structure in the spectral
function  as well as  the anomalous quark dispersion 
relation of the quark can be understood 
in terms of the {\em resonant scattering}
between the incident quark and a particle
near the Fermi surface to make the pairing soft mode.
\end{abstract}

\date{\today}
\maketitle

\section{Introduction}  \label{intro}
The recent data in the RHIC experiment suggest that the 
matter created by the RHIC seems to be a
strongly coupled system with possible quasi-bound hadrons
contained\cite{RHIC,sQGP}.
Furthermore some lattice calculations, though 
in the quenched  approximation, 
are consistent with the possible view that heavy-quark bound systems
such as $J/\psi$ survive the deconfinement transition at
finite temperature $T$\cite{Lattice}.
These developments have a possibility to change drastically the simple
picture of the QGP phase  that the system is composed
of almost free quasi-particles, although
such  nontrivial properties of the QGP phase that it may contain
quasi-hadronic excitations
had been suggested earlier\cite{HK85,DeTar}.

In the present Letter, we shall argue that 
the {\it dense} quark matter at relatively low temperature
can have also non-trivial properties, i.e., a {\em non}-Fermi liquid
ones, if the system is
close to the critical temperature $T_c$ of the 
color-superconducting phase transition \cite{BL84,RW00}
on a rather generic ground.

First we notice 
that the low energy effective models of QCD 
show that the diquark gap at zero temperature may become
as large as $\Delta \sim 100$MeV\cite{ARW98,RSSV98}
at lower densities such as those corresponding to $\mu \simeq 400$MeV
with $\mu$ being the baryon-number chemical potential of the quark.
Accordingly, the ratio $r_{\xi}\equiv\Delta/\mu$, which is
 a measure of the  ratio of the 
inter-particle distance to the pair coherence length,
 may become as large as 0.2 to 0.3.
This value is much larger than those
in the metal superconductors, i.e., $r_{\xi}\simeq 0.001$,
which account for the validity of the mean-field approximation a l\'a
BCS theory  for the electric superconductors.
Thus one sees that large fluctuations of the diquark-pair are expected,
which may invalidate the mean-field approximation,
in the color superconductors\cite{KKKN02,Vosk04}.

In fact, our previous work with the fixed pairing coupling 
showed \cite{KKKN04,KKKN05} that there exists the precursory
soft mode composed of the diquark-pair field even well above $T_c$.
One of the remarkable points there was that 
the coupling  of the quark with the soft mode 
not only modifies the quark  dispersion relation but also
causes a {\em depression} of the quark spectral function 
around the Fermi energy; the depression was so large that
a  ``pseudogap'' is formed  in the density of states (DOS)
of the quark even with the small diquark coupling that was employed.

The above finding
is interesting because they suggest
that the properties of quarks at $T$ above but close to $T_c$
are altered from the typical Fermi-liquid 
to the non-Fermi liquid ones  owing to the coupling to 
the precursory fluctuations of the diquark-pair field.
The mechanism to cause such a 
non-Fermi liquid behavior, however,
 might not have been clarified enough
in the previous work.
The purpose of the present Letter is to elucidate the
mechanism to realize such a non-Fermi liquid behavior
and argue that it is a generic feature 
of the heated quark matter in the vicinity 
of $T_c$ of the color superconductivity.

For this purpose, we examine
what happens in the quark properties in detail
if the pairing fluctuations become stronger by increasing
the diquark coupling constant $G_C$;
we remark that the limiting case can often clarify physics.
Incidentally it may be noticed that  $G_C$ indeed 
could have other values than that in the previous work
 without any contradiction with other principles and phenomenology.
We  calculate the dispersion relation and the spectral function 
of the quark in the T-matrix approximation\cite{KadBay,KKKN04,KKKN05}.

We shall demonstrate the following for the first time:
(1)~ The quark dispersion relation is so largely modified
with the increasing $G_C$ that it becomes seemingly multiple-valued
around the Fermi surface near but above $T_c$ in the strong coupling
regime.\,
(2)~ 
The quark spectral function gets to have
a developed gap-like structure rather than a depression
 near but above $T_c$ for the larger $G_C$.
We shall clarify that  the above non-Fermi liquid
behaviors 
can be nicely understood in terms of 
the {\it resonant scattering}\cite{JML97,HTSC}
between the quark and a particle near the Fermi surface to make the 
pairing soft mode.

\section{Brief summary of formulation}

To describe a system at relatively low $T$ and $\rho$,
it is appropriate to adopt a low-energy effective theory of QCD 
\cite{BR99,SKP99}.
Here, 
we consider the two-flavor quark matter 
in the chiral limit
with the four-Fermi quark-quark interaction\cite{HK94}
\begin{eqnarray}
{\cal L}_C 
= G_C \sum_A ( \bar{\psi} i \gamma_5 \tau_2 \lambda_A C \bar\psi^T )
( \psi^T C i \gamma_5 \tau_2 \lambda_A \psi ),
\label{eqn:lag}
\end{eqnarray}
where $C = i\gamma_2 \gamma_0$ is the charge conjugation operator,
and $\tau_2$ and $\lambda_A$ denote the antisymmetric flavor $SU(2)_f$ 
and color $SU(3)_c$ matrices, respectively.
The quark chemical potentials $\mu$ of each flavor and color 
are taken to be the same\cite{KKKN05}.
We introduce the three momentum cutoff $\Lambda=650$MeV\cite{SKP99}.
As for the diquark coupling constant $G_C$,
we vary it in the range 
$3.11{\rm GeV}^{-2} < G_C < 4.35{\rm GeV}^{-2}$, 
which is the similar range used 
in the references\cite{ARW98,RSSV98,SKP99,BR99}.

In order to see the quasiparticle picture of the quarks,
we examine the spectral function  $\rho_0$ 
and the dispersion relation of the quark $\omega=\omega_-(\bfk)$.
The following formulation is essentially 
a recapitulation of that given in \cite{KKKN05}.

The spectral function $\rho_0$ is given 
by the imaginary part of the retarded 
Green function $G^R( \bfk,\omega )$ as
\begin{eqnarray}
\rho_0( \bfk,\omega ) 
=-\frac1{4\pi} {\rm Tr} [\gamma^0 {\rm Im}G^R ( \bfk,\omega )].
\label{eqn:spectral}
\end{eqnarray}
The Green function $G^R(\bfk,\omega)$ is decomposed into
the quark and anti-quark parts; 
\begin{eqnarray}
G^R(\bfk,\omega)
&=& (G^R_-( \bfk,\omega ) \Lambda_-(\bfk) 
+ G^R_+( \bfk,\omega ) \Lambda_+(\bfk) )\gamma^0
\nonumber \\
&=&
\frac{ \Lambda_-(\bfk) \gamma^0 }{ R_-(\bfk,\omega ) +i\eta }
+ \frac{ \Lambda_+(\bfk) \gamma^0 }{ R_+(\bfk,\omega ) +i\eta },
\label{eqn:G^R}
\end{eqnarray}
with the projection operators
$\Lambda_\mp(\bfk) = ( 1 \pm \gamma^0 \bfgamma\cdot\hat{\bfk})/2$.
The dispersion relations of the quarks and
anti-quarks $ \omega= \omega_\mp( \bfk ) $ are defined by 
the equations 
\begin{eqnarray}
{\rm Re} R_\mp( \bfk,\omega)=0 ,
\label{eqn:ReR}
\end{eqnarray}
respectively.
A remark is in order here: Because $\omega_{\mp}( \bfk ) $ is merely  the
solution of the  real part  but not the whole part of 
the inverse of the Green function $R_{\mp}( \bfk,\omega_{\mp})$,
$\omega_{\mp}( \bfk ) $ may not correspond to the peak position of the
spectral function and hence also may not represent physical
excitations when the imaginary part of the Green function is large.

\begin{figure}
\begin{center}
\includegraphics[width=8.5cm]{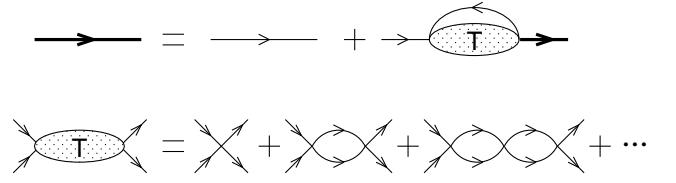} 
\caption{
Feynman diagrams representing the
quark Green function in the 
 T-matrix approximation.
The thin lines represent the free propagator ${\cal G}_0$,
while the bold ones the full propagator ${\cal G}$.
}
\label{fig:Tmat}
\end{center} 
\end{figure}

Our point in the calculation of the quark Green function $ G^R( \bfk,\omega )$
lies in incorporating the diquark-pair fluctuations in the quark self-energy
in the T-matrix approximation\cite{KadBay,KKKN04,KKKN05}
as is diagrammatically shown in Fig.~\ref{fig:Tmat};
all the possible anomalous behaviors of the results will originate
from the fact that the pair fluctuations actually form the soft mode
of the color-superconducting phase transition.

In the Matsubara formalism, the self-energy of quarks 
in the imaginary time $\tilde\Sigma(\bfk,\omega_n)$ is given by
\begin{eqnarray}
\lefteqn{  \tilde\Sigma( \bfp,\omega_n )  }
\nonumber \\
&=& 8
T\sum_m \int \frac{d^3 \bfk }{(2\pi)^3}
\tilde\Xi ( \bfp+\bfk, \omega_n+\omega'_m )
{\cal G}_0 ( \bfk,\omega'_m ),
\label{eqn:Sigma}
\end{eqnarray}
with $\omega_n=(2n+1) \pi T$ being the Matsubara frequency for
fermions and ${\cal G}_0( \bfk,\omega_n ) 
= [ ( i\omega_n + \mu )\gamma^0 - \bfk\cdot \mbox{\boldmath$\gamma$}]^{-1}$
being the free quark propagator.
The T-matrix for the quark-quark scattering $\tilde\Xi(\bfk,\nu_n)$ reads
\begin{eqnarray}
\tilde{\Xi} ( \bfk,\nu_n )
&=& -G_C \frac 1{ 1+G_C {\cal Q}({\bfk},\nu_n) },
\label{eqn:Xi}
\end{eqnarray}
with the lowest order polarization function 
${\cal Q}(\bfk,\nu_n)$\cite{KKKN05} and $\nu_n = 2n\pi T$.
The T-matrix in the real time $\Xi^R( \bfk,\omega ) 
\equiv \tilde\Xi( \bfk,\nu_n )|_{i\nu_n\to \omega+i\eta} $
has all the information about properties of the pair fluctuations.
In particular, the dynamical structure factor
\begin{eqnarray}
S( \bfk,\omega )
&=& -\frac{{\rm Im}~ \Xi^R ( \bfk,\omega )}
{ 2\pi G_C^2 ( 1-{\rm e}^{-\omega/T} ) },
\label{eqn:DSF}
\end{eqnarray}
represents the excitation probability of the pair field
for each $\omega$ and $\bfk$ at finite $T$.
It has 
a prominent peak at the origin near $T_c$ 
according to the softening of the pair fluctuations.

The analytic continuation of $\tilde\Sigma( \bfk,\omega_n )$
to the real axis from the upper-half complex-energy plane
gives the self-energy in the real time
$ \Sigma^R( \bfk,\omega )
= \tilde\Sigma( \bfk,\omega_n )|_{ i\omega_n \to \omega+i\eta }$.
Using the projection operators $\Lambda_\mp(\bfk)$,
the self-energy is decomposed into the quark and anti-quark parts
$\Sigma_{\mp}( \bfk,\omega )
= \Tr [ \Sigma^R( \bfk,\omega ) \Lambda_\mp(\bfk) \gamma^0 ]/2$,
and $R_\mp( \bfk,\omega )$ in Eq.~(\ref{eqn:G^R}) is now found to be
$R_\mp( \bfk,\omega ) 
= \omega +\mu \mp k - \Sigma_\mp( \bfk,\omega )$.
Notice that $\rho_0(\bfk,\omega)$ around the Fermi energy 
is  given almost solely by the quark
part $G^R_-$, and hence 
the corresponding part of 
the self-energy $\Sigma_-(\bfk,\omega)$ is responsible 
for the quasiparticle properties of the quark near the Fermi surface.
For later convenience,
we rewrite  the explicit form of ${\rm Im}\Sigma_-(\bfk,\omega)$ 
given in \cite{KKKN05} slightly 
using $S(\bfk,\omega)$;
\begin{eqnarray}
\lefteqn{ {\rm Im}\Sigma_-(\bfk,\omega) }
\nonumber \\
&=&
\frac{ \pi G_C^2 }2
\int \frac{d^3 \bfq }{(2\pi)^3}
S( \bfk+\bfq , \omega+E_q-\mu ) 
\left\{ 1-\frac{ \hat{\bfk} \cdot \bfq }{ E_q } \right\}
\nonumber \\ &&
\times\left[
( 1 - {\rm e}^{ -( \omega+E_q-\mu )/T } )
\tanh \frac{ E_q-\mu }{2T} \right.
\nonumber \\ && \left.
- ( 1 + {\rm e}^{ - ( \omega+E_q-\mu )/T } ) \right] + (E_q \to -E_q),
\label{eqn:ImS}
\end{eqnarray}
with $E_q = |\bfq|$.

\begin{figure*}
\begin{center}
\includegraphics[width=5.8cm]{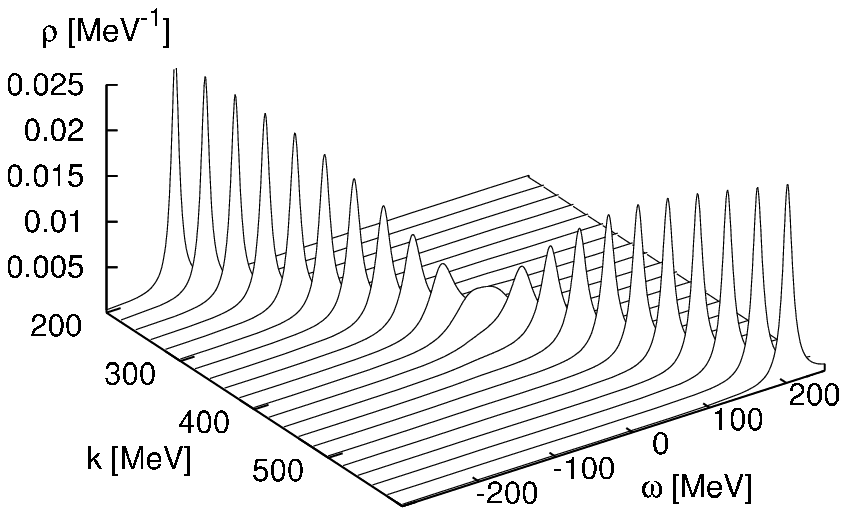}
\includegraphics[width=5.8cm]{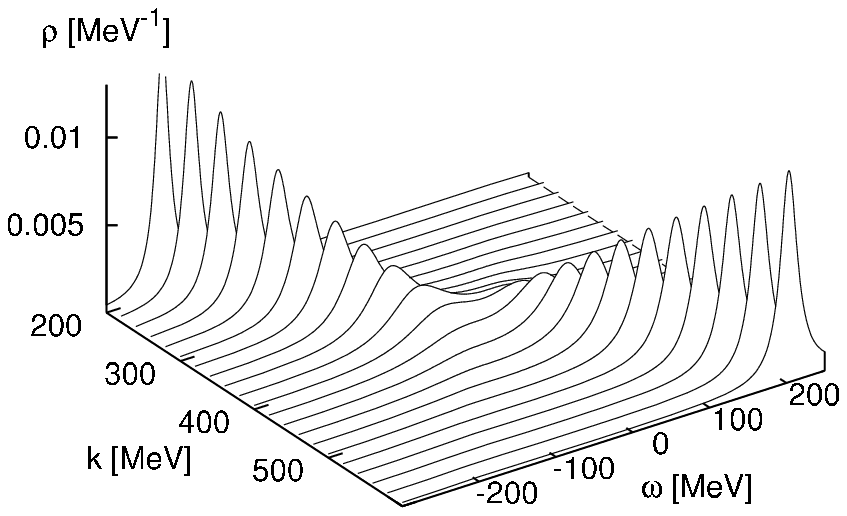}
\includegraphics[width=5.8cm]{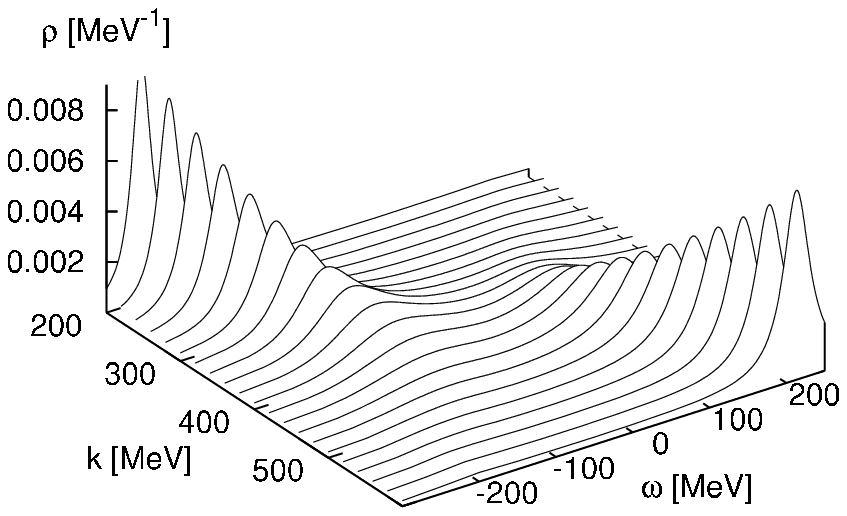} \\
\vspace{5mm}
\includegraphics[width=5.8cm]{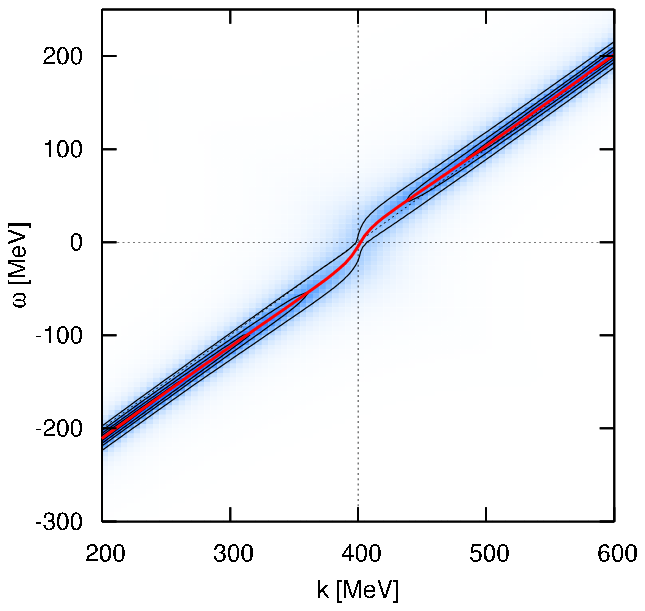} 
\includegraphics[width=5.8cm]{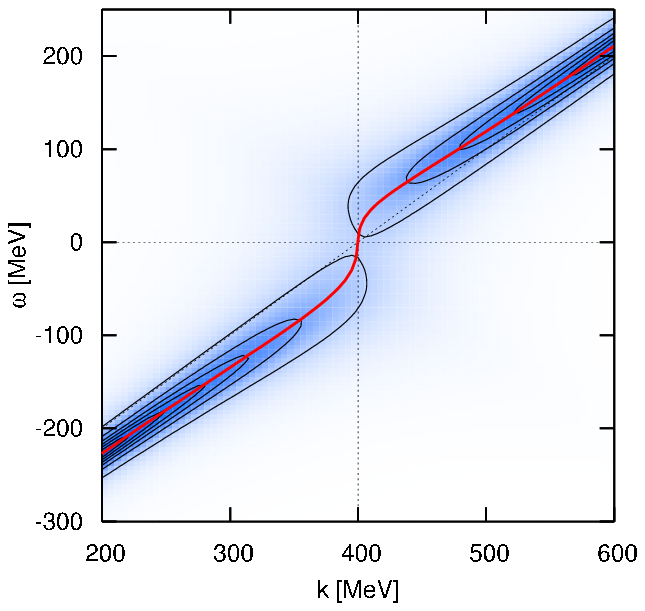} 
\includegraphics[width=5.8cm]{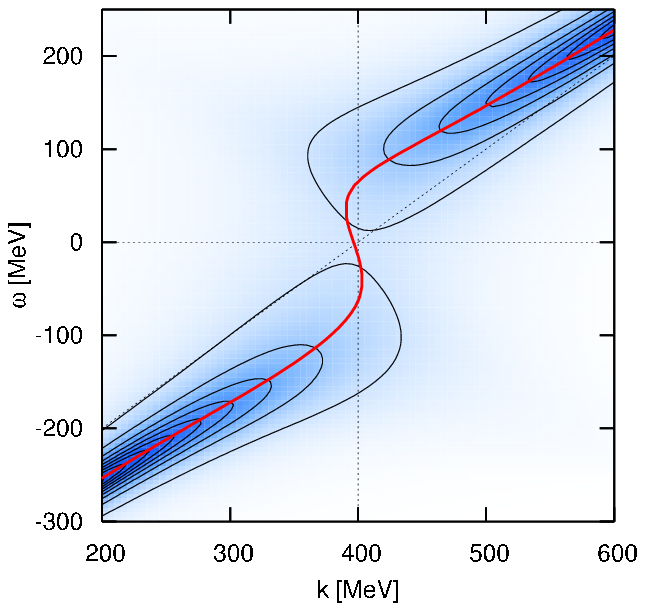} 
\caption{
The spectral function $\rho_0(\bfk,\omega)$ (upper panels)
and the dispersion relation of quarks $\omega=\omega_-(\bfk)$ (lower panel)
for $\mu=400$MeV and $\varepsilon\equiv(T-T_c)/T_c = 0.01$.
In the far left, middle and far right panels,
the diquark coupling constant is taken 
$G_C=3.11{\rm GeV}^{-2}$, 
$3.73 (=3.11\times1.2){\rm GeV}^{-2}$ and 
$4.35 (=3.11\times1.4){\rm GeV}^{-2}$, respectively.
}
\label{fig:spc_map}
\end{center} 
\end{figure*}

\section{Non-fermi liquid behavior and resonant scattering}

In this section, 
we show the spectral function $\rho_0( \bfk,\omega )$ 
and the dispersion relation of quarks $ \omega=\omega_-(\bfk) $
to see the quasiparticle picture of quarks.
As was promised in the Introduction, 
we vary the diquark coupling constant $G_C$
for a fixed $\varepsilon \equiv ( T-T_c )/T_c = 0.01$.

The upper panels of Fig.~\ref{fig:spc_map} 
show the spectral function
$\rho_0( \bfk,\omega )$ for $\mu=400$MeV and $\varepsilon=0.01$ 
with several values of $G_C$, while
the lower panels 
the dispersion relation $ \omega=\omega_-(\bfk) $ together
with the contour map of the strength of $\rho_0( \bfk,\omega )$
represented by the shadow. 

The figures in the far left show 
those with the same $G_C$ ($T_c=40.04$ MeV) as those presented
in Refs.~\cite{KKKN04,KKKN05} for comparison:
Although the detailed properties of them are
given in the previous paper\cite{KKKN05},
the essential points may be summarized
as follows;
(1)~There appears a depression of the quasiparticle peak
in the spectral function $\rho_0(\bfk,\omega)$ 
around the Fermi energy $\omega=0$.
This means that the life-time of the quasiparticles 
near the Fermi surface becomes short
owing to the pair fluctuations.
(2)~The quark dispersion $\omega_-(\bfk)$ shows
a rapid increase around the Fermi momentum.

In the middle and the far right panels of Fig.~\ref{fig:spc_map}, 
$G_C$ is increased $1.2$ and $1.4$ times larger
than that in the far left panels, respectively;
correspondingly, $T_c\rightarrow 59.82$ and $80.11$ MeV.
One sees from the figures that 
as $G_C$ is increased,
the depression of $\rho_0(\bfk,\omega)$
around the Fermi energy becomes more significant, 
and the quasiparticle peaks tend to be so completely depressed
that the profile of $\rho_0(\bfk,\omega)$ 
around the Fermi momentum may be characterized 
more properly with the term of a gap-like structure 
rather than with a depression.
In the lower panels, 
$\omega_-(\bfk)$ around the Fermi-momentum shows more
rapid increase for larger $G_C$, and eventually becomes
triple-valued. Here one should notice that $\omega_-(\bfk)$ 
does not necessarily represent the
real excitation spectrum since it is only 
a zero of the real part of the inverse of the
Green function, 
as is noticed before.
Indeed, one sees only two quasiparticle peaks 
in $\rho_0(\bfk,\omega)$ for $k=k_F$, 
corresponding to the first and third solutions of $\omega_-(\bfk)$.
The reason  of the disappearance of
 the middle solution of $\omega_-(\bfk)$
in $\rho_0(\bfk,\omega)$ will be given later.

\begin{figure}
\includegraphics[width=8.5cm]{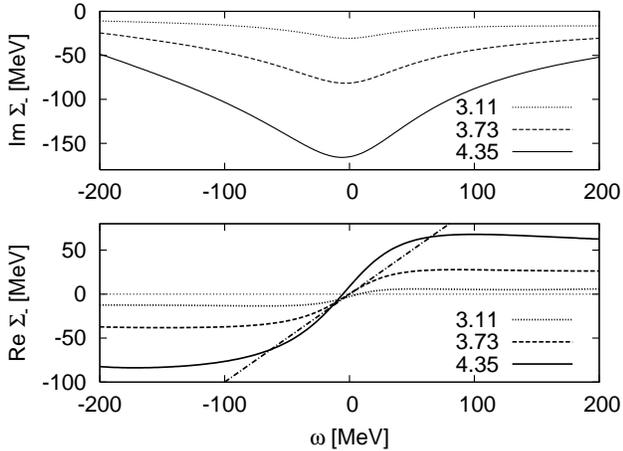} 
\caption{
The imaginary and real parts of the self-energy of quarks
$\Sigma_-( \bfk_F,\omega )$ at $k_F=\mu=400$MeV 
for $G_C=3.11,3.73$ and $4.35$.
As $G_C$ is increased, the peak of Im$\Sigma_-( \bfk_F,\omega )$
as well as the increase of Re$\Sigma_-( \bfk_F,\omega )$ 
around the Fermi energy $\omega=0$ becomes significant.
The dash-dotted line in the bottom panel denotes $ \omega = k-\mu $.
For larger $G_C$, there exist three solutions of 
${\rm Re}~R_-( \bfk,\omega ) = 0$, giving 
the multi-valued dispersion relation.
}
\label{fig:sig_g}
\end{figure}

The above features of $\rho_0(\bfk,\omega)$ and $\omega_-(\bfk)$
can be understood by the behavior of the quark self-energy 
$\Sigma_-( \bfk,\omega )$.
In Fig.~\ref{fig:sig_g}, 
we show the imaginary and real parts of $\Sigma_-( \bfk,\omega )$ 
in the upper and lower panels, respectively, 
at $k=k_F$ for several $G_C$ 
corresponding to each panel of Fig.~\ref{fig:spc_map}.
 For each $G_C$, one sees a peak of $|{\rm Im}\Sigma_-(\bfk,\omega)|$ 
and a rapid increase of Re$\Sigma_-(\bfk,\omega)$
around the Fermi energy $\omega=0$, which is more significant
for larger $G_C$.
In fact, the size of $G_C$ is 
important for the enhancement of $\Sigma_-(\bfk,\omega)$
because of the factor $G_C^2$ in Eq.~(\ref{eqn:ImS})
\footnote{
The dynamical structure factor $S(\bfk,\omega)$ around the origin 
for $\varepsilon=0.01$ is not very sensitive to $G_C$ 
in the range of $G_C$ employed in this work,
while the fluctuations for larger momentum become more significant
as $G_C$ is increased.
}.
We notice that
the behavior of Re$\Sigma_-(\bfk,\omega)$ can be understood
by the growth of $|{\rm Im}\Sigma_-(\bfk,\omega)|$ 
and the Kramers-Kronig relation
\begin{eqnarray}
{\rm Re}\Sigma_-(\bfk,\omega) 
= -\frac1\pi {\rm P}
\int d\omega' {\rm Im}\Sigma_-(\bfk,\omega') / (\omega-\omega').
\label{eqn:DispR}
\end{eqnarray}

The peculiar behavior of $\omega_-(\bfk)$ can be understood as follows.
First, recall that 
 $\omega_-(\bfk)$ at $k=k_F$ is the solution of 
${\rm Re}~R_-( \bfk_F,\omega ) 
= \omega - {\rm Re}~\Sigma_-(\bfk_F,\omega) = 0$.
Accordingly,
 the solutions $\omega_-(\bfk_F)$ are given graphically by the crossing 
points of  Re$\Sigma_-(\bfk_F,\omega)$ and $\omega$ 
denoted by the straight 
dash-dotted line in the lower panel of Fig.~\ref{fig:sig_g}.
One sees how 
there eventually appear the three solutions of $\omega_-(\bfk_F)$
 for large $G_C$, as mentioned before.
A remark is in order here:
The group velocity as defined by $v_g=d\omega_{-}(k)/dk$
is seemingly larger than the speed of light near $\omega_{-}(k)=0$.
However,  it
corresponds to the peak position 
 of the imaginary part of 
the self-energy $|{\rm Im}\Sigma_-(\bfk,\omega)|$, which means
that the excitations around the origin have 
a quite large damping rate, 
and hence there does not appear a peak in the spectral function.
The solution near $\omega=0$ thus does not represent 
 a physical excitation spectrum of the quasiparticle.

\begin{figure}[t]
\includegraphics[width=8.5cm]{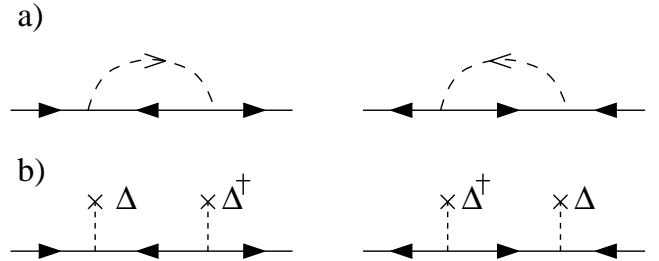} 
\caption{
(a) The self-energy of quarks above $T_c$.
The dashed line denotes the collective mode of 
the pair fluctuations.
The virtual mixing between the particles and holes is induced
by the collective mode.
(b) The self-energy in the BCS theory.
The gap $\Delta$ induces the mixing.
}
\label{fig:mixing}
\end{figure}

\begin{figure}[t]
\includegraphics[width=6cm]{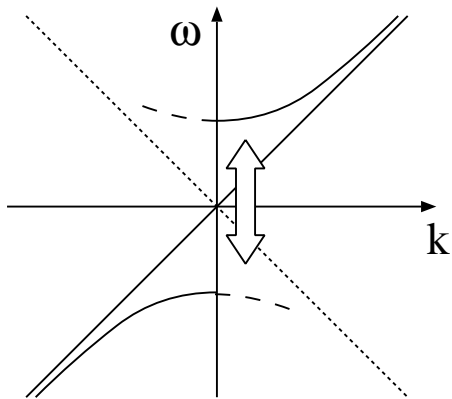} 
\caption{
The quasiparticle energy of quarks.
}
\label{fig:rscatt}
\end{figure}

Now let us discuss the mechanism to realize the gap-like structure 
in the quark spectrum in Fig.~\ref{fig:spc_map}.
Due to the softening of the diquark-pair fluctuations near $T_c$
\cite{KKKN04,KKKN05},
a particle near the Fermi energy is scattered by the soft mode and 
creates a hole, while a hole can 
create a particle by absorbing the soft mode
as shown in Fig.~\ref{fig:mixing}(a).
We remark that the hole must be also 
created near the Fermi surface
because of the energy conservation\cite{KKKN05}.
More intuitively,
the incident particle and a particle  near the Fermi surface,
which may be within the Fermi sea
or  thermally excited one, 
 make a resonant scattering to 
form the pairing soft mode and vice versa.
This resonating processes which are only effective around
the Fermi surface induce a virtual mixing 
between the particles and holes, i.e. a (virtual) Bogoliubov
transformation! 
In other words, because the particle and hole energies
$ \omega = k-\mu $ and $ \omega = \mu-k $ cross at the Fermi energy
as shown in Fig.~\ref{fig:rscatt},
the mixing between the particles and holes leads to 
the level repulsion of the energy spectrum of the particle and hole,
which makes the gap-like structure 
as shown in the right panels of Fig.~\ref{fig:spc_map}.
This mechanism which induces 
the gap-like structure owing to the {\em resonant
scattering} to make the softening pairing mode
is also known in the condensed matter physics \cite{JML97}.

We shall next show how the behavior of  the quark spectrum 
can be nicely reproduced by quantifying the notion of
the resonant scattering.
As was seen above, 
the softening of the pair fluctuations near $T_c$ is responsible
for  the resonant scattering to become effective.
Since the softening is described
as the enhancement of $S( \bfk,\omega )$ around the origin
in the $\bfk$-$\omega$ plane,
 let us approximate the dynamical structure factor
with the delta function at the origin;
\begin{eqnarray}
S( \bfk,\omega ) 
= \Delta^2 \frac{ (2\pi)^4 }{ \pi G_C^2 }
\delta^{(3)}(\bfk) \delta(\omega).
\label{eqn:S}
\end{eqnarray}
The meaning of the 
choice of the pre-factor containing 
a $\Delta$  in front of the delta functions
will be clear later; $\Delta$ is to be related with the imaginary
part of the $T$ matrix $\Xi$ and the boson distribution function
\cite{JML97,HTSC}.
Needless to say, the dynamical structure factor above $T_c$ 
in reality never vanishes in the whole $\omega$-$k$ 
plane nor takes the form of Eq.~(\ref{eqn:S}) even at $T=T_c$;
the fluctuations with finite $\omega$ and $k$ make
a width of the quasiparticles.
However, this simple treatment will be found to clearly illustrate
that the peculiar behavior of the quark spectrum can be nicely
accounted for by the resonant scattering.

A simple calculation with Eq.~(\ref{eqn:ImS}) leads to
\begin{eqnarray}
{\rm Im}\Sigma_- ( \bfk,\omega )
&=& -\pi \Delta^2 \delta( \omega +|\bfk|-\mu ),
\\
{\rm Re}\Sigma_- ( \bfk,\omega )
&=& \Delta^2 \frac1{ \omega +|\bfk|-\mu }.
\end{eqnarray}
Then, $ R_-( \bfk,\omega ) = 0 $ gives real number 
quasiparticle energies
\begin{eqnarray}
\omega_-(\bfk) = \pm \sqrt{ ( |\bfk|-\mu )^2 + \Delta^2 },
\label{eqn:omegaBCS}
\end{eqnarray}
which have the same form as that in the usual BCS theory
with the energy gap $\Delta$;
notice that the quark propagator for the quarks which interact with 
the pairing mode with the dynamical structure factor Eq.~(\ref{eqn:S}) can be 
diagrammatically interpreted as Fig.~\ref{fig:mixing}(b),
i.e., the quarks interacting with a static source of the pair field
$\Delta$.

\section{Summary and concluding remarks}

In this Letter,
motivated with the findings in our previous papers\cite{KKKN04,KKKN05},
we have  tried to elucidate generically that the precursory
pairing soft mode modifies the 
quasi-particle picture of the quark so much that
 the heated quark matter would become a non-Fermi liquid 
in the vicinity of $T_c$ of the color superconductivity.
For this purpose, we have examined the changes of the
quark properties in detail when the diquark coupling
is increased.
We have found that 
in the strong coupling regime the 
quark dispersion relation is modified so greatly
that it  becomes  multiple-valued
around the Fermi surface near but above $T_c$.
The stronger coupling also changes
the quark spectral function as to have a gap-like structure 
rather than a depression.
We have shown that these non-Fermi liquid behaviors can be understood 
in terms of the resonant scattering between 
the incident particle and a particle 
around the Fermi energy to make the pairing soft mode.

Although the present work is based on a model calculation,
the mechanism proposed here 
for realizing a non-Fermi liquid behavior 
of fermion systems  can be model-independent.
This is because the essential ingredient here
is the scattering of a particle near Fermi surface 
to make the soft mode which is 
inherent to any second-order or weak first-order phase transition.
It would be thus intriguing to explore the quasiparticle picture 
near the critical temperature of other phase transitions 
of  QCD matter, including the {\em chiral} phase transition
at finite temperature\cite{KKN}.

We should notice  that there is another mechanism which 
causes  a non-Fermi liquid behavior of ungapped quark matter:
A resummed perturbation theory shows  \cite{rebhan} that 
the coupling of the quark with unscreened long-range gauge (gluonic) fields
gives rise to vanishing of both the residues of the quasiparticle
excitations and the group velocity $v_g=dE/dp$ at the Fermi surface  
at small temperature without recourse to any phase transition.
Notice that the vanishing group velocity implies the infinite
density of states of quasiparticles, which is highly in contrast 
with our result, i.e., the pseudogap formation as a precursory phenomenon
of the CSC at finite $T$.
Although the perturbation theory adopted in \cite{rebhan}
is only applicable for extremely high-density quark matter,
it would be interesting to see how the non-Fermi liquid
behaviors due to unscreened gluonic fields 
 survive and compete with those owing to the soft mode of
CSC at moderate density and temperature.

In this work, we have employed the 
non-selfconsistent T-matrix approximation which is essentially
a linear approximation for the fluctuations.
Nevertheless the results obtained in this  approximation can
be close to reality so long as  $\varepsilon\equiv (T-T_c)/T_c$ 
is not so small as noted in \cite{KKKN04,KKKN05}; see \cite{HTSC}.
In the present work,
the strong fluctuations of the pair-field were induced
with the varied $G_C$ for fixed $\varepsilon = 0.01$.
Similar results  may be obtained
for smaller $\varepsilon$ but with a fixed $G_C$ since
the fluctuations diverge as $\varepsilon$ goes to zero.

The situation, however,  might not become so simple when
the system may enter the Ginzburg region where the nonlinear fluctuation
effects play an essential role and somewhat moderate behavior may be
realized  for the quark properties. 
A renormalization-group treatment, for instance, would be necessary
to incorporate the nonlinear effects of
the fluctuations, which is beyond the scope of the present work and 
left for future investigations.

M. K. is supported by Japan Society for the Promotion of Science for
Young Scientists.
T. Kunihiro is supported by Grant-in-Aide for Scientific Research by
Monbu-Kagaku-sho (No.\ 14540263).
Y. N. is supported by 21st Century COE Program of Nagoya University.

\end{document}